\documentclass[prd,aps,showpacs,showkeys,superscriptaddress,
preprintnumbers,floatfix,byrevtex,twocolumn]{revtex4}

\usepackage{amsbsy}
\usepackage{amsmath}
\usepackage{amssymb}
\usepackage{amsfonts}
\usepackage{graphics}
\usepackage{epsfig}
\usepackage{wasysym}
\usepackage{rotating}
\usepackage{dcolumn}

\newcommand{\half}{\mbox{\small $\frac{1}{2}$}}
\newcommand{\third}{\mbox{\small $\frac{1}{3}$}}
\newcommand{\quarter}{\mbox{\small $\frac{1}{4}$}}
\newcommand{\twelfth}{\mbox{\small $\frac{1}{12}$}}
\newcommand{\Dlr}{\overset{\leftrightarrow}{D}}
\newcommand{\Dl}{\overset{\leftarrow}{D}}
\newcommand{\Dr}{\overset{\rightarrow}{D}}

\begin{document}

\preprint{
\vspace*{7mm}
\hbox{DESY 05-076 }
\hbox{Edinburgh 2005/04 }
\hbox{MPP-2005-52}
}

\title{Investigation of the Second Moment of the Nucleon's
  $g_1$ and $g_2$ Structure Functions in Two-Flavor Lattice QCD}

\author{M.~G\"ockeler}
\affiliation{Max-Planck-Institut f\"ur Physik, F\"ohringer Ring 6,
             80805 M\"unchen, Germany}
\affiliation{Institut f\"ur Theoretische Physik,
             Universit\"at Regensburg, 93040 Regensburg, Germany}
\author{R.~Horsley}
\affiliation{School of Physics, University of Edinburgh,
             Edinburgh EH9 3JZ, UK}
\author{D.~Pleiter}
\affiliation{John von Neumann-Institut f\"ur Computing NIC,
             Deutsches Elektronen-Synchrotron DESY, 15738 Zeuthen, Germany}
\author{P.~E.~L.~Rakow}
\affiliation{Theoretical Physics Division, 
             Department of Mathematical Sciences,
             University of Liverpool, Liverpool L69 3BX, UK}
\author{A.~Sch\"afer}
\affiliation{Institut f\"ur Theoretische Physik,
             Universit\"at Regensburg, 93040 Regensburg, Germany}
\author{G.~Schierholz}
\affiliation{John von Neumann-Institut f\"ur Computing NIC,
             Deutsches Elektronen-Synchrotron DESY, 15738 Zeuthen, Germany}
\affiliation{Deutsches Elektronen-Synchrotron DESY, 22603 Hamburg, Germany}
\author{H.~St\"uben}
\affiliation{Konrad-Zuse-Zentrum f\"ur Informationstechnik Berlin, 
             14195 Berlin, Germany}
\author{J.~M.~Zanotti}
\affiliation{John von Neumann-Institut f\"ur Computing NIC,
             Deutsches Elektronen-Synchrotron DESY, 15738 Zeuthen, Germany}
\collaboration{QCDSF/UKQCD collaboration} \noaffiliation

\date{September 5, 2005}

\begin{abstract}
  The reduced matrix elements $a_2$ and $d_2$ are computed in lattice
  QCD with $N_f=2$ flavors of light dynamical (sea) quarks. For
  proton and neutron targets we obtain as our best estimates
  $d_2^{(p)}=0.004(5)$ and
  $d_2^{(n)}=-0.001(3)$, respectively, in the $\overline{\mbox{MS}}$
  scheme at $Q^2 = 5$ GeV$^2$, while for $a_2$ we find
  $a_2^{(p)}=0.077(12)$ and $a_2^{(n)}=-0.005(5)$, where the errors are 
  purely statistical.
\end{abstract}

\pacs{12.38.Gc, 13.60.Hb, 13.88.+e}

\keywords{Lattice QCD, deep-inelastic scattering, higher twist, 
non-perturbative renormalization}

\maketitle


\section{Introduction}
\label{sec:intro}

The nucleon's second spin-dependent structure function $g_2$ is of
considerable phenomenological interest since at leading order in $Q^2$
it receives contributions from both twist-2 and twist-3 operators.
Consideration of $g_2$ via the operator product expansion (OPE)
\cite{Jaffe} offers the unique possibility of directly assessing
higher-twist effects which go beyond a simple parton model
interpretation.

Neglecting quark masses and contributions of twist greater than two,
one obtains the ``Wandzura-Wilczek'' relation \cite{WW}
\begin{equation}
g_2(x,Q^2) \approx
g_2^{WW}(x,Q^2) = -g_1(x,Q^2) + \int_x^1 \frac{\mbox{d}y}{y}
g_1(y,Q^2)\, ,
\label{eq:g2WW}
\end{equation}
depending only on the nucleon's first spin-dependent structure
function, $g_1(x,Q^2)$. Including mass and gluon dependent terms up to
and including twist-3, $g_2$ can be written \cite{Cortes:1991ja}
\begin{equation}
g_2(x,Q^2) = g_2^{WW}(x,Q^2) + \overline{g_2}(x,Q^2)\, ,
\label{eq:g2}
\end{equation}
where
\begin{equation}
\overline{g_2}(x,Q^2) = -\int_x^1 \frac{\mbox{d}y}{y}
\frac{\mbox{d}}{\mbox{d}y} \left[\frac{m}{M}h_T(y,Q^2) + \xi(y,Q^2)
\right]\, .
\label{eq:g2bar}
\end{equation}
The function $h_T(x,Q^2)$ denotes the transverse polarization density
and has twist two. The contribution from $h_T(x,Q^2)$ to $g_2$ is
suppressed by the quark-to-nucleon mass ratio, $m/M$, and hence is
small for physical up and down quarks. 
The twist-3 term $\xi$ arises from quark-gluon correlations.

From Eqs.~(\ref{eq:g2WW})-(\ref{eq:g2bar}), the moments of $g_2$ are
\begin{eqnarray}
\int_0^1\mbox{d}x\, x^n g_2(x,Q^2) =
\frac{n}{n+1}\left\{ -\int_0^1\mbox{d}x\, x^n g_1(x,Q^2)\right. \nonumber\\
 + \left.\int_0^1\mbox{d}x\, x^{n-1}\bigg[ \frac{m}{M}h_T(x,Q^2) +
\xi(x,Q^2)\bigg] \right\}\, .
\label{eq:mom-g2}
\end{eqnarray}

A leading order OPE analysis with massless quarks shows that the
moments of $g_1$ and $g_2$ are given by \cite{Jaffe}
\begin{eqnarray}
2\int_0^1\mbox{d}x\, x^n g_1(x,Q^2)  
  \!\!&=&\!\! \frac{1}{2} \sum_{f=u,d} e^{(f)}_{1,n}(\mu^2/Q^2,g(\mu))
a_n^{(f)}(\mu)\, , \nonumber\\
&& \label{eq:ope-g1} \\
2\int_0^1\mbox{d}x\, x^n g_2(x,Q^2)  
 \!\!&=&\!\! \frac{1}{2}\frac{n}{n+1} \sum_{f=u,d}
 \big[e^{(f)}_{2,n}(\mu^2/Q^2,g(\mu)) \nonumber\\
\times d_n^{(f)}(\mu) \!\!&-&\!\! e^{(f)}_{1,n}(\mu^2/Q^2,g(\mu))\,
 a_n^{(f)}(\mu)\big]\, ,
\label{eq:ope-g2}
\end{eqnarray}
for even $n \ge 0$ for Eq.~(\ref{eq:ope-g1}) and even $n \ge 2$ for
Eq.~(\ref{eq:ope-g2}), where $f$ runs over the light quark flavors
and $\mu$ denotes the renormalization scale.
The reduced matrix elements $a_n^{(f)}(\mu)$ and $d_n^{(f)}(\mu)$ are
defined by~\cite{Jaffe}
\begin{eqnarray}
\langle \vec{p},\vec{s}| 
 {\cal O}^{5 (f)}_{ \{ \sigma\mu_1\cdots\mu_n \} }
                       | \vec{p},\vec{s} \rangle 
 \!\!&=&\!\! \frac{1}{n+1}a_n^{(f)} \nonumber\\ 
&&\!\! \times [ s_\sigma p_{\mu_1} \cdots p_{\mu_n} 
+ \cdots -\mbox{traces}], \nonumber\\ \label{eq:twist2} \\
\langle \vec{p},\vec{s}| 
 {\cal O}^{5 (f)}_{ [  \sigma \{ \mu_1 ] \cdots \mu_n \} }
                       | \vec{p},\vec{s} \rangle 
   &=& \frac{1}{n+1}d_n^{(f)} \nonumber\\ 
\times [ (s_\sigma p_{\mu_1} \!\!&-&\!\! s_{\mu_1} p_\sigma)
 p_{\mu_2}\cdots p_{\mu_n} 
 + \cdots -\mbox{traces}], \nonumber\\
\label{eq:twist3} 
\end{eqnarray}
\begin{equation}
 {\cal O}^{5 (f)}_{\sigma\mu_1\cdots\mu_n}
   = \left(\frac{\mathrm i}{2}\right)^n\bar{\psi}\gamma_{\sigma} \gamma_5
 \Dlr_{\mu_1}  
 \cdots \Dlr_{\mu_n}
 \psi -\mbox{traces}\, .
\end{equation}
Here $\Dlr=\Dr-\Dl$ and $e_{1,n}^{(f)}$, $e_{2,n}^{(f)}$ are the
Wilson coefficients which depend on the ratio of scales $\mu^2/Q^2$,
the running coupling constant $g(\mu)$ and the quark charges ${\cal
  Q}^{(f)}$,
\begin{equation}
e^{(f)}_{i,n}(\mu^2/Q^2,g(\mu)) = {\cal Q}^{(f)2} \big(1 + 
{\cal O}(g(\mu)^2) \big)\, .
\label{eq:wilsoncoef}
\end{equation}
The symbol$\{\cdots\}$ ($[\cdots]$) indicates symmetrization
(antisymmetrization) of indices. The operator (\ref{eq:twist2}) has twist
two, whereas the operator (\ref{eq:twist3}) has twist three. 
Note that our definitions of $a_2$ and $d_2$ differ by a factor of two
from those in \cite{exp2,exp}.

Using the equations of motion of massless QCD one can rewrite the twist-3
operators
$ {\cal O}^{5 (f)}_{ [  \sigma \{ \mu_1 ] \cdots \mu_n \} } $
such that the dual gluon field strength tensor $\tilde{G}_{\mu \nu}$
and the QCD coupling $g$ appear. 
For $n=2$ one finds
\begin{equation}
{\cal O}^{5 (f)}_{ [  \sigma \{ \mu_1 ] \mu_2 \} } =
- \frac{g}{6} \bar{\psi} \left( \tilde{G}_{\sigma \mu_1} \gamma_{\mu_2}
    + \tilde{G}_{\sigma \mu_2} \gamma_{\mu_1} \right) \psi
  - \mbox{traces} \,,
\end{equation}
so we can define the reduced matrix element $d_2$ in the chiral limit
also by (see, e.g., Ref.~\cite{schaefer})
\begin{eqnarray}
& & - \frac{g}{6} \langle \vec{p},\vec{s}|
 \bar{\psi} \left( \tilde{G}_{\sigma \mu_1} \gamma_{\mu_2}
    + \tilde{G}_{\sigma \mu_2} \gamma_{\mu_1} \right) \psi
  - \mbox{traces}
                       | \vec{p},\vec{s} \rangle \nonumber\\
& & = \frac{1}{3} d_2
[ (s_\sigma p_{\mu_1} - s_{\mu_1} p_\sigma) p_{\mu_2}
 + \cdots -\mbox{traces}] \,.
\end{eqnarray}
This shows (setting $\mu_1 = \mu_2 = 0$) that $d_2$ parametrizes the
magnetic field component of the gluon field strength tensor which is
parallel to the nucleon spin. 
Furthermore we have
\begin{equation}
d_2 = 4\int_0^1\mbox{d}x\, x \xi(x) \ .
\label{eq:d2w2}
\end{equation}
Hence, a calculation of $d_2$ (in the chiral limit) is especially
interesting as it will provide insights into the size of the
quark-gluon correlation term, $\xi(x)$.

The Wilson coefficients (\ref{eq:wilsoncoef}) can be computed
perturbatively, while the reduced matrix elements $a_n^{(f)}$ and
$d_n^{(f)}$ have to be computed non-perturbatively.
In the following we shall drop the flavor indices, unless they are
necessary.

A few years ago we computed the lowest non-trivial moment of $g_2$ in
the quenched approximation~\cite{QCDSF1}. In this paper we give our
results for the reduced matrix elements $a_2$ and $d_2$ in full QCD,
including $N_f=2$ flavors of light dynamical (sea) quarks, using
${\cal O}(a)$-improved Wilson fermions.
We employ the same methods as in the quenched case, in particular the
renormalization of the lattice operators is done entirely
non-perturbatively.


\section{Lattice Operators And Renormalization}
\label{sec:operators}

The lattice calculation divides into two separate tasks. The first
task is to compute the nucleon matrix elements of the appropriate
lattice operators. This was described in detail in~\cite{QCDSF2}.
The second task is to renormalize the operators.  
In the case of multiplicative renormalizability, the renormalized
operator ${\cal O}(\mu)$ is related to the bare operator ${\cal O}(a)$
by
\begin{equation}
   {\cal O}(\mu) = Z_{\cal O}(a\mu)\, {\cal O}(a),
\label{eq:op1}
\end{equation}
where $a$ is the lattice spacing.
In our earlier work~\cite{QCDSF2,QCDSF3}, we computed the renormalization
constants in perturbation theory to one-loop order.
However, this does not account for mixing with lower-dimensional
operators, which we encounter in the case of the reduced matrix
elements $d_n^{(f)}$.
In \cite{QCDSF1} an entirely non-perturbative solution to this problem
was presented for quenched lattice QCD.
Here we shall apply the same approach. We impose the (MOM-like)
renormalization condition~\cite{Martinelli,QCDSF4} (which can also be used
in the continuum)
\begin{equation}
\quarter \,\mbox{Tr} \, \langle q(p)|{\cal O}(\mu)|q(p)\rangle
\Big[\langle q(p)|{\cal O}(a)|q(p)\rangle\, |^{\rm tree}\Big]^{-1}
\underset{p^2 =\mu^2}{=} 1,
\end{equation}
where $|q(p)\rangle$ is a quark state of momentum $p$ in Landau gauge.

In the following we shall restrict ourselves to the case $n = 2$.
Furthermore, we consider quark-line connected diagrams only, as
calculations of quark-line disconnected diagrams are extremely
computationally expensive.
In an attempt to improve on our earlier analysis \cite{QCDSF1}, we
simulate with two non-vanishing values for the nucleon momentum,
$\vec{p}_1 = ( p, 0, 0 )$ and $\vec{p}_2 = ( 0, p, 0 )$,
together with two different polarization directions, described by the
matrices
$\Gamma_1 = \frac{1}{2}(1+\gamma_4)\, {\mathrm i}\gamma_5\gamma_1$ and
$\Gamma_2 = \frac{1}{2}(1+\gamma_4)\, {\mathrm i}\gamma_5\gamma_2$.
Here $p=2\pi/L_S$ denotes the smallest non-zero momentum available on a
periodic lattice of spatial extent $L_S$.
We consider the two combinations $\vec{p}_1$/$\Gamma_2$ and 
$\vec{p}_2$/$\Gamma_1$. For the twist-2 matrix element $a_2$ we use
in both cases the operator
\begin{equation}
{\cal O}^5_{\{214\}} =: {\cal O}^{\{5\}} 
\label{eq:os1}
\end{equation}
as in~\cite{QCDSF1}.

For the twist-3 matrix element $d_2$ we need to use different
operators for our two momentum/polarization combinations. For
$\vec{p}_1$/$\Gamma_2$ and $\vec{p}_2$/$\Gamma_1$ we take
\begin{eqnarray}
{\cal O}^5_{[2\{1] 4\}} \!\!& =&\!\! \third\left(2 {\cal O}^5_{2\{14\}} 
- {\cal O}^5_{1\{24\}} - {\cal O}^5_{4\{12\}}\right) \nonumber \\   
\!\!& =&\!\!\twelfth\bar{\psi}\Big(\gamma_2 \Dlr_{1} \Dlr_{4} + \gamma_2 \Dlr_{4}
  \Dlr_{1} - \half \gamma_1 \Dlr_{2} \Dlr_{4} \nonumber \\
\!\!&- &\!\!\half \gamma_1 \Dlr_{4} \Dlr_{2}
- \half \gamma_4 \Dlr_{1} \Dlr_{2} - \half \gamma_4
  \Dlr_{2} \Dlr_{1}\Big) \gamma_5 \psi \nonumber \\
&=:& {\cal O}^{[5]}_1 \, , \label{eq:o5} \\
{\cal O}^5_{[1\{2] 4\}} \!\!& =&\!\! \third\left(2 {\cal O}^5_{1\{24\}} 
- {\cal O}^5_{2\{14\}} - {\cal O}^5_{4\{21\}}\right) \nonumber \\   
\!\!& =&\!\!\twelfth\bar{\psi}\Big(\gamma_1 \Dlr_{2} \Dlr_{4} + \gamma_1 \Dlr_{4}
  \Dlr_{2} - \half \gamma_2 \Dlr_{1} \Dlr_{4} \nonumber \\
\!\!&- &\!\!\half \gamma_2 \Dlr_{4} \Dlr_{1}
- \half \gamma_4 \Dlr_{2} \Dlr_{1} - \half \gamma_4
  \Dlr_{1} \Dlr_{2}\Big) \gamma_5 \psi \nonumber \\
&=:& {\cal O}^{[5]}_2 \, , \label{eq:o5-2}
\end{eqnarray}
respectively. In the following we shall suppress the index of 
${\cal O}^{[5]}$ unless it is needed.
The operators ${\cal O}^{\{5\}}$ and ${\cal O}^{[5]}$
belong to the representations $\tau_3^{(4)}$ and $\tau_1^{(8)}$,
respectively, of the hypercubic group $H(4)$~\cite{Mandula}. The
operator ${\cal O}^{[5]}$ has dimension five and $C$-parity $+$.
It turns out that there exist two operators of dimension four and
five, respectively, transforming identically under $H(4)$ and having
the same $C$-parity, with which ${\cal O}^{[5]}$ can mix:
\begin{eqnarray}
\twelfth{\mathrm i}\, 
 \bar{\psi} \left(\sigma_{13} \Dlr_{1} -  \sigma_{43} \Dlr_{4}\right)
 \psi \!\!&=:&\!\! {\cal O}^\sigma, \label{eq:osigma} \\
\twelfth \bar{\psi} \left(\gamma_1 \Dlr_{3} \Dlr_{1} - \gamma_1
  \Dlr_{1} \Dlr_{3} \right. \qquad\qquad\quad &&  \nonumber \\ 
\left. -\ \gamma_4 \Dlr_{3}\Dlr_{4} + \gamma_4 \Dlr_{4} \Dlr_{3}\right)
\psi \!\!&=:&\!\! {\cal O}^0 \, , \label{eq:o0}
\end{eqnarray}
for $\vec{p}_1$/$\Gamma_2$, and similarly for $\vec{p}_2$/$\Gamma_1$
with $1\rightarrow 2$.
We use the definition $\sigma_{\mu \nu} = (\mathrm i
/2)[\gamma_\mu,\gamma_\nu]$.

The operator (\ref{eq:o0}) mixes with ${\cal O}^{[5]}$ with a
coefficient of order $g^2$ and vanishes in the tree approximation
between quark states.
We therefore neglect its contribution to the renormalization of ${\cal
  O}^{[5]}$.
The operator ${\cal O}^\sigma$, on the other hand, contributes with a
coefficient $\propto a^{-1}$ and hence must be kept.
We then remain with
\begin{equation}
{\cal O}^{[5]}(\mu) = Z^{[5]}(a\mu) {\cal O}^{[5]}(a) + \frac{1}{a}
Z^\sigma(a\mu) {\cal O}^\sigma(a).
\label{eq:renorm}
\end{equation}
The renormalization constant $Z^{[5]}$ and the mixing coefficient
$Z^\sigma$ are determined from
\begin{eqnarray}
\quarter \,\mbox{Tr} \,\langle q(p)|{\cal O}^{[5]}(\mu)|q(p)\rangle 
\left[\langle   q(p)|{\cal O}^{[5]}(a)|q(p)\rangle\, |^{\rm
    tree}\right]^{-1} 
\hspace*{-5mm}&\underset{p^2 =\mu^2}{=}&\hspace*{-3mm} 1, \nonumber \\
&&\label{eq:cond1} \\
\quarter \,\mbox{Tr} \,\langle q(p)|{\cal O}^{[5]}(\mu)|q(p)\rangle 
\left[\langle
  q(p)|{\cal O}^{\phantom{[}\sigma\phantom{]}}(a)|q(p)\rangle\, |^{\rm
    tree}\right]^{-1} 
\hspace*{-5mm}&\underset{p^2 =\mu^2}{=}&\hspace*{-3mm} 0. \nonumber \\
&&\label{eq:cond2}  
\end{eqnarray}

Rewriting Eq.~(\ref{eq:renorm}) as
\begin{equation}
{\cal O}^{[5]}(\mu) = Z^{[5]}(a\mu)\left( {\cal O}^{[5]}(a) +
  \frac{1}{a} \frac{Z^\sigma(a\mu)}{Z^{[5]}(a\mu)} {\cal
    O}^\sigma(a)\right)\, ,
\label{eq:renorm2}
\end{equation}
we see that ${\cal O}^{[5]}(\mu)$ will have a multiplicative
dependence on $\mu$ only if the ratio $Z^\sigma(a\mu)/Z^{[5]}(a\mu)$
does not depend on $\mu$, which should happen for large enough values
of the renormalization scale. The scale dependence will then completely 
reside in $Z^{[5]}$.

\begin{figure}[t]
  \begin{center}
\vspace*{-2.5cm}
{\includegraphics[width=0.99\hsize]{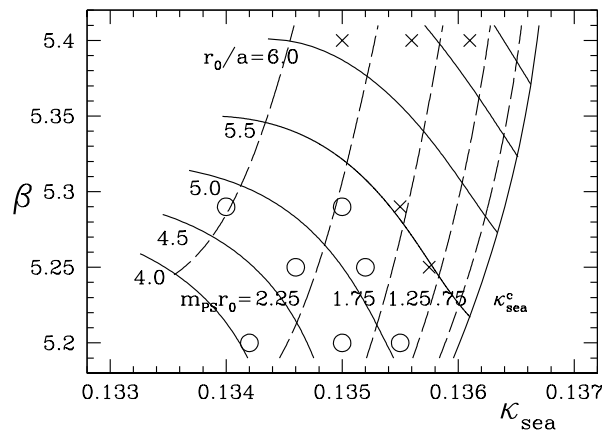}}
\vspace*{-5mm}
\caption{Parameters of our dynamical gauge field configurations, together with
  lines of constant $r_0/a$ (solid lines) and lines of constant
  $m_{\rm PS}\, r_0$ (dashed lines). The simulations are done on
  $24^3\times 48$ ($\times$) and $16^3\times 32$ ($\Circle$) lattices,
  respectively.}
    \label{fig:parameters}
\end{center}
\end{figure}


\section{Simulation Details}
\label{sec:details}

To reduce cut-off effects, we use non-perturbatively $O(a)$ improved
Wilson fermions. The calculation is done at four different values of
the coupling, $\beta$, and at three different sea quark masses each.
The latter are specified by the hopping parameter $\kappa_{\rm sea}$.
We use the force parameter $r_0$ to set the scale, with $r_0 = 0.467$
fm. Our lattice spacings range from $a=0.07$ to $0.09$ fm.
The actual parameters, as well as the corresponding values of $r_0/a$ and
the pseudoscalar meson masses, are given in 
Table~\ref{table:parameters} and shown pictorially in
Fig.~\ref{fig:parameters}.

\begin{table}[tb]
\caption{Lattice parameters:
 gauge coupling $\beta$, sea quark hopping parameter $\kappa_{\rm sea}$,
 lattice volume, number of trajectories, $r_0/a$ and
 pseudoscalar meson mass.}
  \label{table:parameters}
 \begin{ruledtabular}
 \begin{tabular}{cccccl}
        $\beta$ & $\kappa_{\rm sea}$ & Volume & $N_{\rm traj}$ & $r_0/a$ & 
        \multicolumn{1}{c}{$m_{\rm PS} a$}     \\ \hline
    5.20 & 0.13420 & $16^3\times 32$ & O(5000) & 4.077(70) & 0.5847(12)    \\
    5.20 & 0.13500 & $16^3\times 32$ & O(8000) & 4.754(45) & 0.4148(13)    \\
    5.20 & 0.13550 & $16^3\times 32$ & O(8000) & 5.041(53) & 0.2907(15)    \\
    5.25 & 0.13460 & $16^3\times 32$ & O(5800) & 4.737(50) & 0.4932(10)    \\
    5.25 & 0.13520 & $16^3\times 32$ & O(8000) & 5.138(55) & 0.3821(13)    \\
    5.25 & 0.13575 & $24^3\times 48$ & O(5900) & 5.532(40) & 0.25638(70)   \\
    5.29 & 0.13400 & $16^3\times 32$ & O(4000) & 4.813(82) & 0.5767(11)    \\
    5.29 & 0.13500 & $16^3\times 32$ & O(5600) & 5.227(75) & 0.42057(92)   \\
    5.29 & 0.13550 & $24^3\times 48$ & O(2000) & 5.566(64) & 0.32688(70)   \\
    5.40 & 0.13500 & $24^3\times 48$ & O(3700) & 6.092(67) & 0.40301(43)   \\
    5.40 & 0.13560 & $24^3\times 48$ & O(3500) & 6.381(53) & 0.31232(67)   \\
    5.40 & 0.13610 & $24^3\times 48$ & O(3500) & 6.714(64) & 0.22120(80)
\end{tabular}
\end{ruledtabular}
\end{table}

The quark matrix elements for the renormalization constants are computed
using a momentum source \cite{QCDSF4}. 
Performing the Fourier transform at the source suppresses the effect
of fluctuations:
The statistical error in this case is $\propto(VN_{\rm conf})^{-1/2}$
for $N_{\rm conf}$ configurations on a lattice of volume $V$,
resulting in small statistical uncertainties even for a small number
of configurations, at least five in our case.
Hence, the main source of statistical uncertainty in our final results
is from the calculation of the bare matrix elements, not the $Z$
values.

Nucleon matrix elements are determined from the ratio of three-point
to two-point correlation functions
\begin{equation}
\label{eq:ratio}
{\cal R}(t,\tau;\vec{p};{\cal O})\, =
 \frac{C_\Gamma (t,\tau;\vec{p},{\cal O})} 
        {C_2(t,\vec{p})} \ ,
\end{equation}
where $C_2$ is the unpolarized baryon two-point function with a source
at time 0 and sink at time $t$, while the three-point
function $C_\Gamma$ has an operator ${\cal O}$ insertion at time
$\tau$.
To improve our signal for non-zero momentum we average over both
polarization/momentum combinations.

Correlation functions are calculated on configurations
taken at a distance of 5-10 trajectories using 4-8 different locations of
the fermion source. We use binning to obtain an effective distance of
20 trajectories. The size of the bins has little effect on the
error, which indicates auto-correlations are small.


\section{Computation of Renormalization Constants}
\label{sec:renorm}

The twist-2 operator defined in Eq.~(\ref{eq:os1}) is renormalized
multiplicatively with the renormalization factor $Z^{\{5\}}(a\mu)$,
while the renormalization of the twist-3 operators in
Eqs.~(\ref{eq:o5}), (\ref{eq:o5-2}) is more complicated
due to the mixing effects described in Section~\ref{sec:operators}.
Since the renormalization of ${\cal O}^{[5]}_1$ and ${\cal O}^{[5]}_2$
is identical (up to lattice artefacts) we consider only 
${\cal O}^{[5]}_1$.

The calculation of the non-perturbative renormalization factors is a
non-trivial exercise, the full details of which are beyond the scope
of this paper.
Here we restrict ourselves to a short outline of the procedure.
More details can be found in Section~5.2.3 of Ref.~\cite{timid}, and
a fuller account will be given in a forthcoming publication.

\begin{figure}[t]
  \begin{center}
{\includegraphics[width=0.98\hsize]{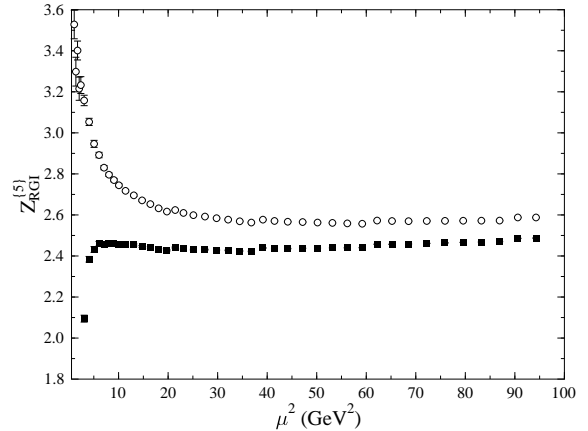}}
\caption{$Z^{\{5\}}_{\rm RGI}$ calculated in the $\overline{\rm MS}$ scheme
  (circles) and in a MOM scheme (filled squares) at $\beta=5.40$.
  The scale is fixed using $r_0=0.467$fm.  }
    \label{fig:Za2}
\end{center}
\end{figure}

Firstly, a chiral extrapolation of the non-perturbative
renormalization factors is performed at fixed $\beta$ and fixed momentum.
The extrapolation is performed linearly in $(r_0 m_{\rm PS})^2 =
((r_0/a)am_{\rm PS})^2$, where for each value of $\beta$ we use the
chirally extrapolated value of $r_0/a$ (see Table~3 of
Ref.~\cite{Gockeler:2005rv}).
We then apply continuum perturbation theory to calculate the
renormalization group invariant renormalization factor $Z_{\rm
RGI}$ from the chirally extrapolated $Z$s \cite{timid}.
This can be done in various schemes, e.g., the $\overline{\rm MS}$
scheme, and should lead for any scheme to the same
momentum-independent value of $Z_{\rm RGI}$, at least for sufficiently
large momenta.  For this step, we use $r_0\Lambda_{\overline{\rm
    MS}}=0.617$~\cite{Gockeler:2005rv}.
In Fig.~\ref{fig:Za2}, we show the $\mu$-dependence of
$Z^{\{5\}}_{\rm  RGI}$ computed in the $\overline{\rm MS}$ scheme
and in a continuum MOM scheme at $\beta=5.40$. While in both cases
a reasonable plateau appears, the plateau values do not coincide
exactly, and we take the difference as a measure of the uncertainty
of our $Z$s, caused by our incomplete knowledge of the perturbative 
expansion.

The final step requires $Z_{\rm RGI}$ to be converted to
$Z_{\overline{\rm MS}}$ at some renormalization scale,
which is done perturbatively, and the result depends on the value of
$\Lambda_{\overline{\rm MS}}$ in physical units. From
$r_0\Lambda_{\overline{\rm MS}}=0.617$ and $r_0 = 0.467$ fm we obtain
$\Lambda_{\overline{\rm MS}} = 261$ MeV.

As mentioned above, the renormalization of the twist-3 operator in
Eqs.~(\ref{eq:o5}), (\ref{eq:o5-2}) has further complications
due to the mixing effects described in Section~\ref{sec:operators}.
In this case it is unclear how to convert our MOM results to the
$\overline{\mbox{MS}}$ scheme. So we shall stick to the MOM numbers.
For the comparison of our results with experimental determinations
this does not cause problems, because no QCD corrections have been
taken into account in the analysis of the experiments and hence
different schemes are not distinguished.

\begin{figure}[t]
  \begin{center}
{\includegraphics[width=0.98\hsize]{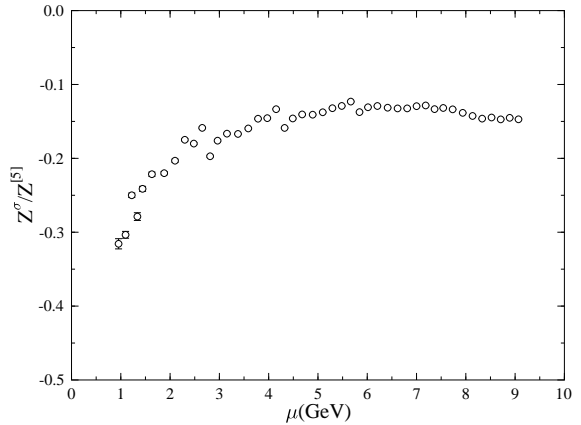}}
\caption{The ratio $Z^{\sigma}(a\mu)/Z^{[5]}(a\mu)$ at $\beta = 5.40$}
    \label{fig:Zd2}
\end{center}
\end{figure}

In Fig.~\ref{fig:Zd2} we plot the ratio $Z^\sigma(a\mu)/Z^{[5]}(a\mu)$
as a function of $\mu$ for $\beta = 5.40$. As expected, a plateau
develops for larger values of $\mu$, and therefore the operator
${\cal O}^{[5]}(\mu)$ only depends on $\mu$ multiplicatively.

\section{Results for Reduced Matrix Elements}
\label{sec:ME}

In order to compute the reduced matrix elements in
Eqs.~(\ref{eq:twist2}) and (\ref{eq:twist3}), we calculate the ratio
of three- to two-point correlation functions ${\cal R}$, as given in
Eq.~(\ref{eq:ratio}), for the operators defined in
Eqs.~(\ref{eq:os1})-(\ref{eq:o0}).
The bare operator matrix elements are obtained from the ratio ${\cal
  R}$ by
\begin{equation}
{\cal R}_{a_2} = \frac{1}{2\kappa_{\rm sea}}\frac{1}{6} M\, p\, a_2\, ,\ 
{\cal R}_{d_2} = \frac{1}{2\kappa_{\rm sea}}\frac{1}{3} M\, p\, d_2\, .
\end{equation}
We define the continuum quark fields by $\sqrt{2\kappa_{\rm sea}}$ 
times the lattice quark fields.
The factor for ${\cal R}_{d_2}$ is the same for all three operators 
${\cal O}^{[5]}$, ${\cal O}^{\sigma}$ and ${\cal O}^0$.

In Tables~\ref{table:bareME1} and \ref{table:bareME2} 
we present our results for the bare matrix
elements of the operators ${\cal O}^{\{5\}}$, ${\cal O}^{[5]}$, ${\cal
  O}^{\sigma}$ and ${\cal O}^0$ defined in
Eqs.~(\ref{eq:os1})-(\ref{eq:o0}) for $u$ and $d$ quarks in the proton.

\begin{table*}
\caption{Bare (unrenormalized) matrix elements $a_2$, $d_2^{[5]}$,
  for $u$ and $d$ quarks in the proton for our entire set of 
  $(\beta,\,\kappa_{\rm sea})$ combinations.}
\label{table:bareME1}
\begin{ruledtabular}
\begin{tabular}{ccdddddddd}
$\beta$ & $\kappa_{\rm sea}$ & \multicolumn{1}{c}{$a_2^{(u)}$}
& \multicolumn{1}{c}{$a_2^{(d)}$} & \multicolumn{1}{c}{$d_2^{[5]\,(u)}$}
& \multicolumn{1}{c}{$d_2^{[5]\,(d)}$} 
\\ \hline  \hline 
  5.20 & 0.13420 & 0.142(18) & -0.0318(78) & -0.0143(23) & 0.0005(14) \\
  5.20 & 0.13500 & 0.123(22) & -0.032(11) & -0.0329(59) & 0.0094(35) \\
  5.20 & 0.13550 & 0.131(32) & -0.061(22) & -0.057(14) & 0.0064(59) 
\\ \hline
  5.25 & 0.13460 & 0.113(12) & -0.0389(51) & -0.0165(25) & 0.0023(13) \\
  5.25 & 0.13520 & 0.110(19) & -0.0281(74) & -0.0310(39) & 0.0069(17) \\
  5.25 & 0.13575 & 0.1107(74) & -0.0345(47) & -0.0575(28) & 0.0074(15) 
\\ \hline
  5.29 & 0.13400 & 0.1141(77) & -0.0255(35) & -0.0033(11) & -0.00009(63) \\
  5.29 & 0.13500 & 0.0989(90) & -0.0281(45) & -0.0252(19) & 0.0046(11) \\
  5.29 & 0.13550 & 0.1228(65) & -0.0302(26) & -0.0468(23) & 0.00783(92)
\\ \hline
  5.40 & 0.13500 & 0.1195(44) & -0.0227(24) & -0.02135(99) & 0.00232(61) \\
  5.40 & 0.13560 & 0.1238(63) & -0.0331(34) & -0.0445(26) & 0.0069(11) \\
  5.40 & 0.13610 & 0.127(13) & -0.0277(60) & -0.0674(48) & 0.0103(25) 
\end{tabular}
\end{ruledtabular}
\end{table*}

\begin{table*}
\caption{Bare (unrenormalized) matrix elements 
  $d_2^{\sigma}/a$ and $d_2^{0}$ for $u$ and $d$ quarks in the proton
  for our entire set of $(\beta,\,\kappa_{\rm sea})$ combinations.}
\label{table:bareME2}
\begin{ruledtabular}
\begin{tabular}{ccdddddddd}
$\beta$ & $\kappa_{\rm sea}$ & \multicolumn{1}{c}{$d_2^{\sigma\,(u)}/a$}
& \multicolumn{1}{c}{$d_2^{\sigma\,(d)}/a$} 
& \multicolumn{1}{c}{$d_2^{0\,(u)}$}
& \multicolumn{1}{c}{$d_2^{0\,(d)}$} 
\\ \hline  \hline 
  5.20 & 0.13420 &  -0.220(19) & 0.046(8) & -0.0312(46) & 0.0096(22) \\
  5.20 & 0.13500 &  -0.305(29) & 0.077(13) & -0.039(10) & 0.0145(49) \\
  5.20 & 0.13550 &  -0.395(60) & 0.080(21) & -0.063(14) & 0.0194(75) 
\\ \hline
  5.25 & 0.13460 &  -0.252(17) & 0.045(6) & -0.0371(34) & 0.0150(28) \\
  5.25 & 0.13520 &  -0.239(23) & 0.063(10) & -0.0329(61) & 0.0131(42) \\
  5.25 & 0.13575 &  -0.353(13) & 0.0638(44) & -0.0463(39) & 0.0141(20) 
\\ \hline
  5.29 & 0.13400 &  -0.213(9) & 0.0379(35) & -0.0322(23) & 0.0086(12) \\
  5.29 & 0.13500 &  -0.258(13) & 0.0518(42) & -0.0312(34) & 0.0118(21) \\
  5.29 & 0.13550 &  -0.338(10) & 0.0651(36) & -0.0390(25) & 0.0120(13) 
\\ \hline
  5.40 & 0.13500 &  -0.301(8) & 0.0595(33) & -0.0396(18) & 0.01231(84) \\
  5.40 & 0.13560 &  -0.385(15) & 0.0723(50) & -0.0502(26)  & 0.0137(15) \\
  5.40 & 0.13610 &  -0.420(25) & 0.087(9) & -0.0411(60) & 0.0178(39) 
\end{tabular}
\end{ruledtabular}
\end{table*}

The corresponding renormalized (reduced) matrix elements for the 
renormalization scale $\mu^2 = 5 \, \mbox{GeV}^2$ are given in
Tables~\ref{table:renME1} and \ref{table:renME2}.
While the superscripts $(u)$ and $(d)$ again 
refer to $u$ and $d$ quarks in the proton, the matrix elements for 
proton and neutron targets are denoted by $(p)$ and $(n)$, respectively.
For $a_2$ the latter are given by
\begin{eqnarray}
a_2^{(p)} &=& {\cal Q}^{(u)\,2} a_2^{(u)} +  
              {\cal Q}^{(d)\,2} a_2^{(d)} , \\  
a_2^{(n)} &=& {\cal Q}^{(d)\,2} a_2^{(u)} +  
              {\cal Q}^{(u)\,2} a_2^{(d)} 
\end{eqnarray}
and similarly for $d_2$. The renormalized values of $d_2^{(f)}$ 
for $f=u,d$ in the proton are calculated from
\begin{equation}
d_2^{(f)} = Z^{[5]} d_2^{[5](f)} + 
 \frac{1}{a} Z^\sigma d_2^{\sigma(f)} \, .
\end{equation}

In the lines for $\kappa_{\rm sea} = \kappa_c$,
Tables~\ref{table:renME1} and \ref{table:renME2} contain results in the
chiral limit, obtained by an extrapolation linear in $(r_0 m_{\rm PS})^2$.
The scale has been fixed from the value of $r_0/a$ at the respective 
quark masses using $r_0 = 0.467 \, \mbox{fm}$. Alternatively, we could 
have worked with the chirally extrapolated values of $r_0/a$. This would
increase $d_2^{(p)}$ and $d_2^{(u)}$ by up to twice the statistical error
but would leave the other observables almost unaffected.
On the other hand, setting $r_0 = 0.5 \, \mbox{fm}$ or varying 
$r_0\Lambda_{\overline{\rm MS}}$ between 0.572 and 0.662 (corresponding
to the combined statistical and systematic errors given in 
Ref.~\cite{Gockeler:2005rv}) leads only to rather small changes in the
final results. 

\begin{figure}[t]
  \begin{center}
{\includegraphics[width=0.99\hsize]{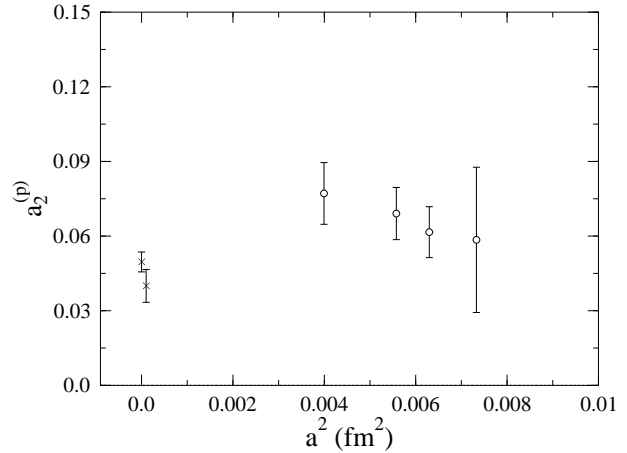}}
\caption{The chirally extrapolated reduced matrix element $a_2$ 
  for the proton target renormalized at
  the scale $\mu^2 \equiv Q^2 = 5$ GeV$^2$ as a function of the 
  lattice spacing $a$. The crosses denote phenomenological determinations.}
    \label{fig:a2p}
\end{center}
\end{figure}

\begin{figure}[t]
  \begin{center}
{\includegraphics[width=0.99\hsize]{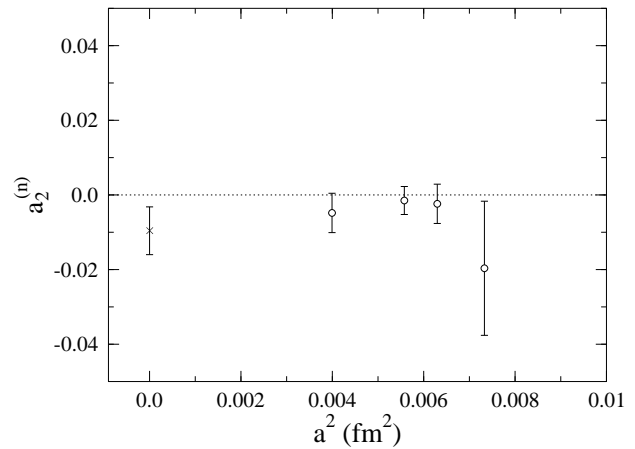}}
\caption{The chirally extrapolated reduced matrix element $a_2$ 
  for the neutron target 
  renormalized at the scale $\mu^2 \equiv Q^2 = 5$ GeV$^2$ as a function of 
  the lattice spacing $a$. The cross denotes the phenomenological 
  value.}
    \label{fig:a2n}
\end{center}
\end{figure}

Let us first focus on the results for the twist-2 matrix element
$a_2$. In Fig.~\ref{fig:a2p} we show the chirally extrapolated 
renormalized results for $a_2$ in the proton in the 
$\overline{\mbox{MS}}$ scheme as a function of the lattice spacing $a$.
It should however be noted that the data at $\beta = 5.20$, i.e., those for
the largest lattice spacing are to be considered with caution, 
because potentially they are affected by lattice artefacts.
For $a_2$ the dependence on the quark mass turns out to be rather small.
On the other hand, we do not attempt a continuum extrapolation of
the chirally extrapolated results. Instead we take the value at 
our smallest lattice spacing ($\beta = 5.4$) as our best estimate:
$a_2^{(p)}=0.077(12)$. This is consistent with earlier quenched 
results~\cite{QCDSF1}, indicating that quenching effects are small.

At the physical pion mass, we compare with two results taken from the
literature which are obtained from an analysis of experimental
data. 
The larger value is taken from an earlier analysis performed by Abe
{\it et al.} \cite{exp2}, while the lower point is extracted from a recent
analysis by Osipenko {\it et al.} \cite{Osipenko:2005nx} with the help
of the perturbative Wilson coefficient. In the $\overline{\mbox{MS}}$ 
scheme with anticommuting $\gamma_5$, we use the two-loop expression for the
Wilson coefficient described in Ref.~\cite{Zijlstra:1993sh}.
To avoid large logarithms, we set $Q^2=\mu^2= 5$ GeV$^2$ to obtain
\begin{equation}
\label{eq:wilson}
e^{(f)}_{1,2} = {\cal Q}^{(f)2} \times 1.03075\, .
\end{equation}

We do not see exact agreement between our chirally extrapolated value
and those obtained from experimental data, but there are still several
sources of systematic error in our final number.
Firstly, our simulation only involves the calculation of connected
quark diagrams. That is, we do not consider the (computationally
expensive) case where an operator couples to a disconnected quark
loop, although such disconnected diagrams are not expected to
contribute in the large $x$ region.
Secondly, our results are restricted to the heavy pion world,
$m_{\rm PS}>550$~MeV. In this region we observe a linear dependence of
our results on $m_{\rm PS}^2$.
A more advanced functional form guided by chiral perturbation theory,
such as those proposed for the moments of unpolarized nucleon
structure functions \cite{Detmold} or nucleon magnetic moments
\cite{chiral}, may be required.
One such form has been suggested in \cite{Detmold:2002nf}, but only
for iso-vector matrix elements.
So we attempt to gain an estimate of the systematic uncertainty due to
our linear extrapolation by comparing results for $a_2^{(u-d)}$ in the
chiral limit using both a linear extrapolation and the form proposed
in \cite{Detmold:2002nf}
\begin{eqnarray}
a_2^{(u-d)}(m_\pi^2) &=& a_2^{(u-d)} \left( 1+c_{\rm LNA}\, m_\pi^2
  \log\frac{m_\pi^2}{m_\pi^2 + \mu^2} \right) \nonumber \\
&&\qquad + b_2\frac{m_\pi^2}{m_\pi^2 + m_b^2}\ ,
\label{eq:chextrap}
\end{eqnarray}
where the authors recommend a preferred value for the LNA coefficient
as $c_{\rm LNA} = -(0.48 g_A^2 + 1)/(4\pi f_\pi)^2$ and $b_2$ is
constrained by the heavy quark limit to be
\begin{equation}
b_2^{(u-d)} = \frac{5}{27} - a_2^{(u-d)} (1-\mu^2 c_{\rm LNA} )\ .
\end{equation}
We set $\mu=0.25$~GeV as proposed in \cite{Detmold:2002nf} and find
at $\beta=5.29$, $a_2^{(u-d)} = 0.214(29)$ employing a linear
extrapolation and $a_2^{(u-d)} = 0.183(9)$ using
Eq.~(\ref{eq:chextrap}), suggesting there is a $15\%$ systematic error
in our linear extrapolation.

Finally, we have not considered finite size effects
\cite{Detmold:2005pt} in this work, and our data do not yet allow us
to perform a decent continuum extrapolation.

\begin{table*}
\caption{Renormalized matrix elements for the renormalization scale 
$\mu^2 = 5 \, \mbox{GeV}^2$ in the $\overline{\mbox{MS}}$ scheme. 
The superscripts $(u)$ and $(d)$ refer to $u$ and $d$ quarks in the proton.}
\label{table:renME1}
\begin{ruledtabular}
\begin{tabular}{ccdddd}
$\beta$ & $\kappa_{\rm sea}$ & \multicolumn{1}{c}{$a_2^{(u)}$}
& \multicolumn{1}{c}{$a_2^{(d)}$} & \multicolumn{1}{c}{$d_2^{(u)}$}
& \multicolumn{1}{c}{$d_2^{(d)}$} 
\\ \hline  \hline 
  5.20 & 0.13420 & 0.194(27) & -0.044(11) &  0.0360(59) & -0.0113(29) \\
  5.20 & 0.13500 & 0.168(32) & -0.044(15) &  0.039(12) & -0.0082(65) \\
  5.20 & 0.13550 & 0.179(45) & -0.083(30) &  0.034(28) & -0.015(11) \\
  5.20 & $\kappa_c$ & 0.154(65) & -0.079(37) &  0.040(31) & -0.011(14) 
\\ \hline
  5.25 & 0.13460 & 0.154(19) & -0.0532(76) &  0.0335(53) & -0.0070(24) \\
  5.25 & 0.13520 & 0.150(27) & -0.038(10) &  0.0109(79) & -0.0047(34) \\
  5.25 & 0.13575 & 0.151(13) & -0.0472(70) &  0.0024(54) & -0.0050(25) \\
  5.25 & $\kappa_c$ & 0.149(24) & -0.042(12) & -0.0169(89) & -0.0036(41) 
\\ \hline
  5.29 & 0.13400 & 0.159(14) & -0.0356(53) &  0.0468(27) & -0.0094(13) \\
  5.29 & 0.13500 & 0.138(15) & -0.0392(67) &  0.0284(43) & -0.0064(20) \\
  5.29 & 0.13550 & 0.171(13) & -0.0421(43) &  0.0201(44) & -0.0056(17) \\
  5.29 & $\kappa_c$ & 0.167(24) & -0.0469(84) & -0.0008(70) & -0.0026(28) 
\\ \hline
  5.40 & 0.13500 & 0.170(12) & -0.0323(39) &  0.0499(27) & -0.0127(13) \\
  5.40 & 0.13560 & 0.176(13) & -0.0471(55) &  0.0401(57) & -0.0097(22) \\
  5.40 & 0.13610 & 0.181(21) & -0.0394(88) &  0.019(10) & -0.0094(46) \\
  5.40 & $\kappa_c$ & 0.187(28) & -0.056(11) &  0.010(12) & -0.0056(50) 
\end{tabular}
\end{ruledtabular}
\end{table*}

\begin{table*}
\caption{Renormalized matrix elements for the renormalization scale 
$\mu^2 = 5 \, \mbox{GeV}^2$ in the $\overline{\mbox{MS}}$ scheme. 
The superscripts $(p)$ and $(n)$ denote the matrix elements for proton 
and neutron targets, respectively.}
\label{table:renME2}
\begin{ruledtabular}
\begin{tabular}{ccdddd}
$\beta$ & $\kappa_{\rm sea}$ & \multicolumn{1}{c}{$a_2^{(p)}$}
& \multicolumn{1}{c}{$a_2^{(n)}$} & \multicolumn{1}{c}{$d_2^{(p)}$}
& \multicolumn{1}{c}{$d_2^{(n)}$}
\\ \hline  \hline 
  5.20 & 0.13420 & 0.081(12) & 0.0022(55) &  0.0148(26) & -0.0010(14) \\
  5.20 & 0.13500 &   0.070(14) & -0.0008(75) &  0.0166(55) & 0.0008(32) \\
  5.20 & 0.13550 &   0.070(20) & -0.017(14) &  0.013(13) & -0.0028(58) \\
  5.20 & $\kappa_c$ & 0.058(29) & -0.020(18) &  0.017(14) & -0.0002(71) 
\\ \hline
  5.25 & 0.13460 &   0.0627(82) & -0.0065(36) &  0.0141(24) & 0.0006(12) \\
  5.25 & 0.13520 &  0.063(12) & -0.0004(53) &  0.0043(36) & -0.0009(18) \\
  5.25 & 0.13575 &  0.0620(58) & -0.0041(31) &  0.0005(24) & -0.0019(13) \\
  5.25 & $\kappa_c$ & 0.062(10) & -0.0024(53) & -0.0079(40) & -0.0035(21) 
\\ \hline
  5.29 & 0.13400 &  0.0668(61) & 0.0019(25) &  0.0198(12) & 0.00105(64) \\
  5.29 & 0.13500 &  0.0570(65) & -0.0021(31) &  0.0119(19) & 0.00031(99) \\
  5.29 & 0.13550 &  0.0715(57) & 0.0003(19) &  0.0083(20) & -0.00028(89) \\
  5.29 & $\kappa_c$ & 0.069(10) & -0.0015(38) & -0.0006(31) & -0.0012(15) 
\\ \hline
  5.40 & 0.13500 & 0.0720(50) & 0.0045(17) &  0.0208(12) & -0.00009(63) \\
  5.40 & 0.13560 &  0.0731(58) & -0.0014(24) &  0.0168(25)  & 0.0001(11) \\
  5.40 & 0.13610 &  0.0760(93) & 0.0026(43) &  0.0072(46) & -0.0021(23) \\
  5.40 & $\kappa_c$ & 0.077(12) & -0.0048(53) &  0.0039(54) & -0.0013(26) 
\end{tabular}
\end{ruledtabular}
\end{table*}

Our results for $a_2$ in the neutron are shown in Fig.~\ref{fig:a2n}.
They are hardly different from zero. Taking again the value for 
$\beta = 5.4$ as our best estimate, we end up with $a_2^{(n)}=-0.005(5)$,
in agreement with the result from the analysis of Abe
{\it et al.}  \cite{exp2}.

From $a_2^{(p)}$ and $a_2^{(n)}$ in the chiral limit we calculate (see
Eq.~(\ref{eq:ope-g1})) the second moment of the polarized structure
function $g_1$ for the proton and neutron.
Using the Wilson coefficient given in Eq.~(\ref{eq:wilson}) we find
\begin{eqnarray}
\int_0^1\mbox{d}x\, x^2 g_1^p(x,Q^2) \!\!&=&\!\! \frac{1.03075}{4}
a_2^p = 0.0170(18) \, , \\
\int_0^1\mbox{d}x\, x^2 g_1^n(x,Q^2) \!\!&=&\!\! \frac{1.03075}{4}
a_2^p = -0.0013(8) .
\end{eqnarray}

\begin{figure}[t]
  \begin{center}
{\includegraphics[width=0.99\hsize]{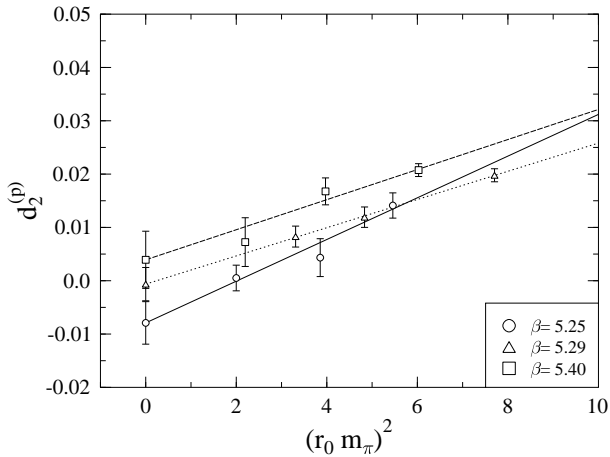}}
\caption{The chiral extrapolation of the reduced matrix element $d_2$ 
  for the proton target renormalized at
  the scale $\mu^2 \equiv Q^2 = 5$ GeV$^2$.}
    \label{fig:d2pc}
\end{center}
\end{figure}

\begin{figure}[t]
  \begin{center}
{\includegraphics[width=0.99\hsize]{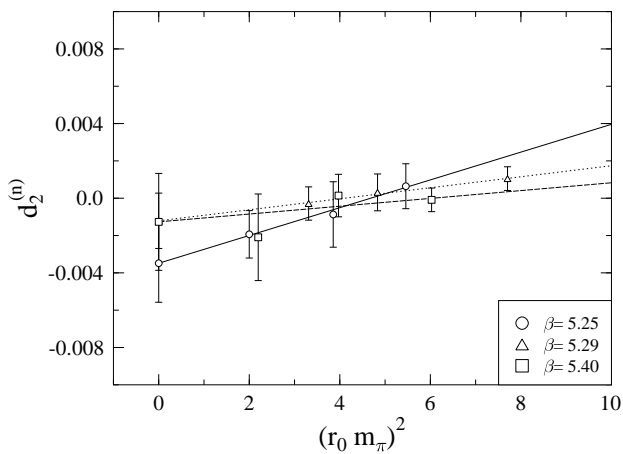}}
\caption{The chiral extrapolation of the reduced matrix element $d_2$ 
  for the neutron target renormalized at
  the scale $\mu^2 \equiv Q^2 = 5$ GeV$^2$.}
    \label{fig:d2nc}
\end{center}
\end{figure}

\begin{figure}[t]
  \begin{center}
{\includegraphics[width=0.99\hsize]{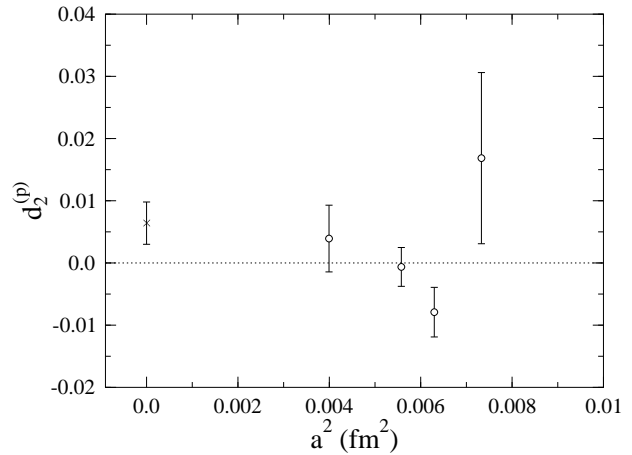}}
\caption{The chirally extrapolated reduced matrix element $d_2$ 
  for the proton target renormalized at
  the scale $\mu^2 \equiv Q^2 = 5$ GeV$^2$ as a function of the 
  lattice spacing $a$. The cross denotes the phenomenological value.}
    \label{fig:d2p}
\end{center}
\end{figure}

\begin{figure}[t]
  \begin{center}
{\includegraphics[width=0.99\hsize]{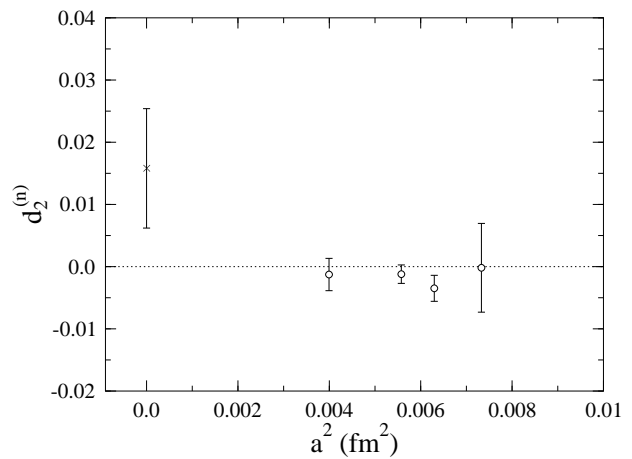}}
\caption{The chirally extrapolated reduced matrix element $d_2$ 
  for the neutron target renormalized
  at the scale $\mu^2 \equiv Q^2 = 5$ GeV$^2$ as a function of the
  lattice spacing $a$. The cross denotes the phenomenological value.}
    \label{fig:d2n}
\end{center}
\end{figure}

We now turn our attention to the second moment of $g_2$.
We find that our data for $d_2$ also exhibit a linear behavior in
$m_{\rm PS}^2$. 
While this is not unexpected at the large pion masses where our
simulations are performed, this linear behavior will not necessarily
continue near the chiral limit.
Unfortunately, the dependence of $d_2$ on the pion mass
near the chiral limit is not yet known. 
Therefore in this work we perform only a linear extrapolation of $d_2$
to the chiral limit.
In Figs.~\ref{fig:d2pc} and \ref{fig:d2nc} we plot some of the data
versus $(r_0 m_{\rm PS})^2$ together with the linear extrapolations. 
The chirally extrapolated results for $d_2$ in the proton and neutron
are shown in Figs.~\ref{fig:d2p} and \ref{fig:d2n}, respectively.
At our smallest lattice spacing we obtain in the chiral limit
\begin{eqnarray}
d_2^{(p)} &=& \phantom{-} 0.004(5), \\
d_2^{(n)} &=& - 0.001(3).
\end{eqnarray}
The errors are statistical only. Taking the behavior of $a_2^{(u-d)}$
as a guide, the chiral extrapolation might introduce a $15\%$
systematic uncertainty.
For $d_2^{(p)}$ the other systematic uncertainties discussed above
would amount to an additional error of about 0.005, while $d_2^{(n)}$
is almost unaffected.
Our result for the proton agrees very well with the experimental
number~\cite{exp}, while for the neutron the experimental result
differs from ours by two standard deviations.
A more precise experimental value would be most desirable in case of
the neutron.

From Eq.~(\ref{eq:mom-g2}), the moments of $g_2$ receive contributions
from $g_1$ and $\overline{g_2}$, the second of which contains a mass
dependent term and a gluon insertion dependent term.
From Eq.~(\ref{eq:g2bar}), the second moment of $\overline{g_2}$ is
(dropping the explicit $Q^2$ dependence)
\begin{equation}
\frac{1}{6}d_2 = 
\int_0^1\mbox{d}x\, x^2 \overline{g_2}(x) =
\int_0^1\mbox{d}x\, x \frac{2}{3}\bigg[ \frac{m}{M}h_T(x) +
\xi(x)\bigg] \, ,
\label{eq:mom2-g2bar}
\end{equation}
so if $d_2$ vanishes in the chiral limit, then 
$\int_0^1 \mathrm d x x \xi (x) $ must also vanish.
Our results lead us to conclude that for the $n=2$ moment the
Wandzura-Wilczek relation~\cite{WW}
\begin{equation}
\int_0^1 \mbox{d}x\, x^2 g_2(x,Q^2)
= - \frac{2}{3} \int_0^1 \mbox{d}x\, x^2 g_1(x,Q^2)
\end{equation}
is satisfied within errors for both proton and neutron targets.

From the expression in Eq.~(\ref{eq:g2bar}), we also expect the first
moment of $\overline{g_2}$ to vanish in the chiral limit.
Combining these two observations with the Burkhardt-Cottingham sum
rule \cite{Burkhardt:1970ti}, $\int_0^1 g_2(x) \mbox{d}x=0$, and the
knowledge that from elastic scattering processes $g_2$
receives non-trivial higher-twist contributions at $x=1$ (see, for
example, Eqs.~(4), (5) of \cite{Osipenko:2005nx}),
we expect that there should be some sort of smooth
transition at intermediate $x$, which presents an interesting
challenge for the planned experiments at JLab~\cite{JLab}.


\section{Conclusions}
\label{sec:conclusions}

We have calculated the second moments of the proton and neutron's
spin-dependent $g_1$ and $g_2$ structure functions in lattice QCD with
two flavors of ${\cal O}(a)$-improved Wilson fermions.
A key feature of our investigation is the use of non-perturbative
renormalization and the inclusion of operator mixing in our extraction
of the twist-2 and twist-3 matrix elements.

Our result for $a_2^{(p)}=0.077(12)$ for the proton is somewhat larger
than what follows from analyses of experimental data, while for the 
corresponding result for the neutron, we find a small but negative value,
$a_2^{(n)}=-0.005(5)$, in agreement with experiment.
Note that the errors are purely statistical and do not include any 
systematic uncertainties, although we estimate a systematic
uncertainty of approximately $15\%$ arising from the chiral
extrapolation.

For the twist-3 matrix element, $d_2$, our results agree very well
with experiment and are consistent with zero, leading us to the
conclusion that higher-twist effects occur only at large or intermediate
$x$.


\section*{Acknowledgments}

J.Z. would like to thank W.~Detmold for useful discussions regarding
the chiral extrapolation of $a_2$.
The numerical calculations have been done on the Hitachi SR8000 at LRZ
(Munich), on the Cray T3E at EPCC (Edinburgh) under PPARC grant
PPA/G/S/1998/00777 \cite{UKQCD} and on the APE{\it 1000} at DESY
(Zeuthen). We thank the operating staff for support. This work was
supported in part by the DFG (Forschergruppe
Gitter-Hadronen-Ph\"anomenologie) and by the EU Integrated
Infrastructure Initiative Hadron Physics (I3HP) under contract
RII3-CT-2004-506078.


\begin{thebibliography}{99}
\bibitem{Jaffe}
  R.~L.~Jaffe,
  Comments Nucl.\ Part.\ Phys.\  {\bf 19}, 239 (1990);
  R.~L.~Jaffe and X.~D.~Ji,
  Phys.\ Rev.\ D {\bf 43}, 724 (1991);
  J.~Blumlein and N.~Kochelev,
  Nucl.\ Phys.\ B {\bf 498}, 285 (1997)
  [arXiv:hep-ph/9612318].

\bibitem{WW}
  S.~Wandzura and F.~Wilczek,
  Phys.\ Lett.\ B {\bf 72}, 195 (1977).

\bibitem{Cortes:1991ja}
  J.~L.~Cortes, B.~Pire and J.~P.~Ralston,
  Z.\ Phys.\ C {\bf 55}, 409 (1992).

\bibitem{exp2}
  K.~Abe {\it et al.}  [E143 collaboration],
  Phys.\ Rev.\ D {\bf 58}, 112003 (1998)
  [arXiv:hep-ph/9802357].

\bibitem{exp}
  P.~L.~Anthony {\it et al.}  [E155 Collaboration],
  Phys.\ Lett.\ B {\bf 553}, 18 (2003)
  [arXiv:hep-ex/0204028].

\bibitem{schaefer}
  B.~Ehrnsperger, L.~Mankiewicz and A.~Sch\"afer,
  Phys.\ Lett.\ B {\bf 323}, 439 (1994)
  [arXiv:hep-ph/9311285].

\bibitem{QCDSF1}
  M.~G\"ockeler {\it et al.},
  Phys.\ Rev.\ D {\bf 63}, 074506 (2001)
  [arXiv:hep-lat/0011091].

\bibitem{QCDSF2}
  M.~G\"ockeler {\it et al.}, 
  Phys.\ Rev.\ D {\bf 53}, 2317 (1996)
  [arXiv:hep-lat/9508004].

\bibitem{QCDSF3}
  M.~G\"ockeler {\it et al.}, 
  Nucl.\ Phys.\ B {\bf 472}, 309 (1996)
  [arXiv:hep-lat/9603006].

\bibitem{Martinelli}
  G.~Martinelli, C.~Pittori, C.~T.~Sachrajda, M.~Testa and A.~Vladikas,
  Nucl.\ Phys.\ B {\bf 445}, 81 (1995)
  [arXiv:hep-lat/9411010].

\bibitem{QCDSF4}
  M.~G\"ockeler {\it et al.},
  Nucl.\ Phys.\ B {\bf 544}, 699 (1999)
  [arXiv:hep-lat/9807044].

\bibitem{Mandula}
  M.~Baake, B.~Gem\"unden and R.~Oedingen,
  J.\ Math.\ Phys.\  {\bf 23}, 944 (1982)
  [Erratum-ibid.\  {\bf 23}, 2595 (1982)];
  J.~E.~Mandula, G.~Zweig and J.~Govaerts,
  Nucl.\ Phys.\ B {\bf 228}, 109 (1983).

\bibitem{timid}
  M.~G\"ockeler, R.~Horsley, D.~Pleiter, P.~E.~L.~Rakow and G.~Schierholz
                  [QCDSF Collaboration],
  arXiv:hep-ph/0410187.

\bibitem{Gockeler:2005rv}
  M.~G\"ockeler {\it et al.}, 
  arXiv:hep-ph/0502212.

\bibitem{Osipenko:2005nx}
  M.~Osipenko {\it et al.},
  Phys.\ Rev.\ D {\bf 71}, 054007 (2005)
  [arXiv:hep-ph/0503018].

\bibitem{Detmold}
  W.~Detmold {\it et al.},
  %
  Phys.\ Rev.\ Lett.\  {\bf 87}, 172001 (2001)
  [arXiv:hep-lat/0103006];
  J.~W.~Chen and X.~Ji,
  %
  Phys.\ Rev.\ Lett.\  {\bf 87}, 152002 (2001)
  [Erratum-ibid.\  {\bf 88}, 249901 (2002)]
  [arXiv:hep-ph/0107158];
  D.~Arndt and M.~J.~Savage,
  %
  Nucl.\ Phys.\ A {\bf 697}, 429 (2002)
  [arXiv:nucl-th/0105045].

\bibitem{chiral}
  M.~G\"ockeler {\it et al.},
  [QCDSF Collaboration],
  Phys.\ Rev.\ D {\bf 71}, 034508 (2005)
  [arXiv:hep-lat/0303019];
  D.~B.~Leinweber, D.~H.~Lu and A.~W.~Thomas,
  %
  Phys.\ Rev.\ D {\bf 60}, 034014 (1999)
  [arXiv:hep-lat/9810005];
  E.~J.~Hackett-Jones, D.~B.~Leinweber and A.~W.~Thomas,
  %
  Phys.\ Lett.\ B {\bf 489}, 143 (2000)
  [arXiv:hep-lat/0004006];
  T.~R.~Hemmert and W.~Weise,
  %
  Eur.\ Phys.\ J.\ A {\bf 15}, 487 (2002)
  [arXiv:hep-lat/0204005];
  R.~D.~Young, D.~B.~Leinweber and A.~W.~Thomas,
  Phys.\ Rev.\ D {\bf 71}, 014001 (2005)
  [arXiv:hep-lat/0406001].

\bibitem{Detmold:2002nf}
  W.~Detmold, W.~Melnitchouk and A.~W.~Thomas,
  Phys.\ Rev.\ D {\bf 66}, 054501 (2002)
  [arXiv:hep-lat/0206001].

\bibitem{Detmold:2005pt}
  W.~Detmold and C.~J.~Lin,
  Phys.\ Rev.\ D {\bf 71}, 054510 (2005)
  [arXiv:hep-lat/0501007].

\bibitem{Zijlstra:1993sh}
  E.~B.~Zijlstra and W.~L.~van Neerven,
  Nucl.\ Phys.\ B {\bf 417}, 61 (1994)
  [Erratum-ibid.\ B {\bf 426}, 245 (1994)].

\bibitem{Burkhardt:1970ti}
  H.~Burkhardt and W.~N.~Cottingham,
  Annals Phys.\  {\bf 56} (1970) 453.

\bibitem{JLab}
 Z.E. Meziani, private communication.

\bibitem{UKQCD}
  C.~R.~Allton {\it et al.}  [UKQCD Collaboration],
  Phys.\ Rev.\ D {\bf 65}, 054502 (2002)
  [arXiv:hep-lat/0107021].


\end{thebibliography}
\end{document}